\documentclass[pre,twocolumn,showpacs,twoside,floatfix]{revtex4}
\usepackage{graphicx,amsmath,amssymb,mathptm}
\begin{document}
\title{Synchronization transition 
  in scale-free networks: Clusters of synchrony}
\author{Deok-Sun Lee}
\affiliation{Theoretische Physik, Universit\"{a}t des Saarlandes,
66041 Saarbr\"{u}cken, Germany}
\date{\today}
\begin{abstract}
We study the synchronization transition  
in scale-free networks that display 
power-law asymptotic behaviors in their degree distributions. 
The critical coupling strength 
and the order-parameter critical exponent 
derived by the mean field approach depend on 
the degree exponent $\lambda$, which implies a 
close connection between structural organization and the emergence of 
 dynamical order in complex systems.
We also derive the finite-size scaling behavior of the order parameter 
finding that the giant cluster of synchronized nodes 
is formed in different ways between 
scale-free networks with $2<\lambda<3$ and those with $\lambda>3$.
\end{abstract}
\maketitle

\section{Introduction}
Much attention has recently been paid to 
integrative approaches to complex systems     
in diverse fields.  
Graph-theoretic 
analyses on social, biological, and informational 
networks consisting of 
relevant elements (nodes) and interactions (edges)   
have found common features such as  
short average distance between nodes, 
high clustering,  and 
scale-free (SF) topology, which means that 
the degree distribution is 
of a power-law form~\cite{watts98,albert02r}. 
These structural characteristics 
have been found to be evidence of 
competitive evolution processes~\cite{barabasi99}   
as well as ground for the robustness of complex systems 
against external attacks or internal errors~\cite{albert00}. 
However, the relation of the structural organization to 
dynamical complexity remains to be explored. 
Recent works on 
epidemic spreading~\cite{satorras01}, avalanche 
dynamics~\cite{watts02,goh03}, and reaction-diffusion 
processes~\cite{gallos04,catanzaro04} in complex networks 
identified the significance of network topology 
in dynamic behavior. 
Here we focus on collective synchronization 
phenomena that appear in various physical and biological systems 
including Josephson junction arrays, cardiac pacemaker cells, 
flashing fireflies, and  brains~\cite{strogatz01} 
as well as are 
required in controlling artificial systems~\cite{korniss03}.
While each element of these systems 
can be described simply by a limit-cycle 
oscillator, the coupling topologies are not always regular.  
For instance, the neuronal network with synaptic coupling 
in human brains turns out to have SF topology~\cite{eguiluz03,shin04},  
whose synchrony is essential in the recognition of  
spatio-temporal patterns~\cite{hopfield01}. 
Also in parallel simulations where tasks are distributed among many 
processing elements, 
weak random couplings among them 
can prevent the spread of virtual times from diverging~\cite{korniss03}.
Our interest is in how the emergence of 
dynamical order is intertwined with structural complexity.

The linear stability analysis of  
the synchronized state in complex networks  of oscillators
has shown that the synchronizability, quantified by the 
ratio of the largest and the second smallest eigenvalues 
of the Laplacian matrix, would be suppressed 
by a few ``center'' nodes being overloaded by 
the traffic of communication~\cite{barahona02,nishikawa02,hong04}. 
However, the effects of topological features on 
the phase transition from the desynchronized state 
to the synchronized state demand further investigation. 
In small-world (SW) networks~\cite{watts98}, 
the synchronization transition 
is observed at a finite critical coupling strength and 
associated critical exponents  
are shown to be equal to those of 
the  globally-coupled case~\cite{kuramoto,strogatz00,daido} 
meaning both belong to the same universality class~\cite{hong02}. 
In this paper we report on the synchronization transition 
in SF  networks of limit-cycle oscillators.
The degree distribution of the SF networks takes a 
power-law form $p_d(k)\sim k^{-\lambda}$ for $k\gg 1$. 
We find by the mean-field approach that the critical coupling strength is 
much smaller than those of completely random graphs~\cite{er61} or 
SW networks~\cite{watts98} that have Poisson degree distributions, 
and the order-parameter critical exponent varies continuously 
with the degree exponent $\lambda$ as long as $2<\lambda<5$, 
illustrating the relation between the structural organization and 
the emergence of   dynamical order. 
We also show how the mean-field approach enables to compute 
the order parameter in the critical regime for finite system size
and derive the finite-size scaling behavior of the order parameter 
finding different nature  of the synchronization transition 
between SF networks with $2<\lambda<3$ and $\lambda>3$. 
Our numerical data confirm the analytic results.

The paper is organized as follows. The model for the coupled limit-cycle 
oscillators on a network is introduced and the order parameter is defined 
in the following section. In Sec.~\ref{sec:sc}, we use the mean-field 
approach to set up self-consistent equations  
and find the critical point and order-parameter critical exponent.
The finite-size scaling behavior is derived in Sec.~\ref{sec:fss}.  
We summarize and discuss our work in Sec.~\ref{sec:summary}.
  
\section{Model and order parameter}
\label{sec:model}
We consider $N$ coupled limit-cycle oscillators whose phases 
$\{\phi_i(t)|i=1,2,\ldots,N\}$ 
evolve according to
\begin{equation}
{d\phi_i \over dt} = \omega_i - \frac{J}{\langle k\rangle} 
\sum_{j=1}^N e_{ij}\sin (\phi_i -\phi_j),
\label{eq:coupled}
\end{equation}
where $J$ is the coupling strength, 
$e_{ij}=1 (0)$ if oscillators $i$ and $j$ are 
connected (disconnected), and 
$\langle k\rangle$ denotes 
the average degree  
$\sum_i k_i/N$ with $k_i = \sum_j e_{ij}$.
The natural frequencies $\omega_i$'s are distributed 
following the Gaussian probability 
distribution $g(\omega_i)=(2\pi)^{-1/2}
e^{-\omega_i^2/2}$.  
A system of globally-coupled oscillators evolving 
by Eq.~(\ref{eq:coupled}), 
i.e.,  the case of $e_{ij}=1$ for all $i\ne j$ and $\langle k\rangle=N-1$, 
defines the Kuramoto model~\cite{kuramoto} and has been used 
as an exactly solvable model of collective synchronization~\cite{strogatz01}.
Here we are interested in   sparse SF coupling networks with 
$\langle k\rangle = \mathcal{O}(1)$ and $\lambda>2$. 

\begin{figure}
\includegraphics[width=\columnwidth]{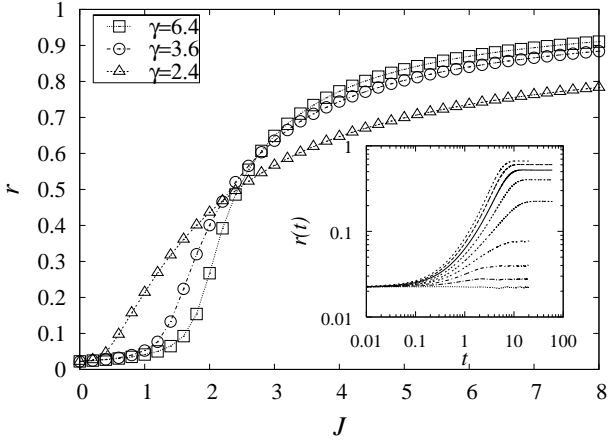}
\caption{Order parameter $r=\lim_{t\to\infty} r(t)$
  versus the coupling strength $J$ for 
$N=1600$, $\langle k\rangle=4$, and $\lambda=6.4$, $3.6$, and $2.4$. 
Inset: Time evolution of $r(t)$ for $\lambda=3.6$ 
with $J=0.0, 0.4, 0.8, 1.2, 1.6, 2.0, 2.4, 2.8$ and $3.2$ from bottom 
to top.}
\label{fig:r}
\end{figure}

Synchronization may be quantified by the 
amplitude order parameter $r(t)$ defined by  
\begin{equation}
r(t)\, e^{i\theta(t)}=
\frac{1}{N}\sum_{j=1}^N e^{i\phi_j(t)}.
\end{equation}
The average phase $\theta(t)$ 
satisfies $d\theta(t)/dt =\bar{\omega}$ with  
$\bar{\omega}=(1/N)\sum_{i=1}^N \omega_i$ 
for a given frequency realization $\{\omega_i\}$. 
To compute $r(t)$, 
we performed a numerical integration of Eq.~(\ref{eq:coupled}) 
on SF networks 
that are generated using the static model introduced in Ref.~\cite{goh01}. 
$r(t)$ is 
averaged over several hundreds of different 
frequency and network realizations, $\{\omega_i\}$ and $\{e_{ij}\}$.
As shown in the inset of Fig.~\ref{fig:r}, $r(t)$ 
saturates to a value $r$ in the long-time limit.
We plot $r$ for $N=1600$ as a function of $J$ in Fig.~\ref{fig:r}, 
evaluated  by $r=(2/T)\sum_{t=T/2+1}^T r(t)$, 
where we set $T$ so as to guarantee 
that the saturation of $r$ has already begun at time $T/2$.
The numerical results imply that the transition from 
the desynchronized state ($r=0$) 
to the synchronized state ($r$ non-zero) 
in the thermodynamic limit $N\to\infty$ 
exhibits a dependence on the coupling topology characterized 
by the degree exponent.
The critical phenomena related to the synchronization transition in 
SF networks can be understood by the mean-field approach. 

\section{Self-consistent equations and critical exponent}
\label{sec:sc}
In this section, we obtain the mean-field solution to Eq.~(\ref{eq:coupled}) 
and find the critical point and order-parameter critical exponent.
Assuming the  same magnitude of effective field 
between each pair of coupled oscillators, 
one obtains the one-body differential equation 
from Eq.~(\ref{eq:coupled}) as 
$d\phi_i /dt=
\omega_i-(J/\langle k\rangle) k_i \bar{r}(t) \sin \phi_i$, 
where the effective field $\bar{r}(t)$ satisfies 
\begin{equation}
\bar{r}(t) e^{i\bar{\theta}(t)}=
\frac{\sum_{i=1}^N \sum_{j=1}^N e_{ij}\exp(i\phi_j)}
{\sum_{i=1}^N \sum_{j=1}^N e_{ij}}, 
\end{equation}
and the relation 
$\bar{\theta}(t)\to 0$ in the thermodynamic limit is used.
$\bar{r}(t)$ will be computed in a self-consistent way. 

In the stationary state,  
the oscillators are either {\it locked} or {\it drifting}~\cite{kuramoto,strogatz00}. 
For given $\bar{r}=\lim_{t\to\infty} \bar{r}(t)$, 
{\it locked} oscillators have their frequencies 
in the region $|\omega_i|\leq Jk_i \bar{r}/\langle k\rangle$ and thus 
their phases are locked at $\sin^{-1}(\omega_i\langle k\rangle/(Jk_i\bar{r}))$
while 
the frequencies of {\it drifting} oscillators 
are in the region $|\omega_i|>Jk_i \bar{r}/\langle k\rangle$ and thus 
$d\phi_i/dt\ne 0$. 
{\it Drifting} oscillators do not contribute to the order parameter 
due to the symmetry of the frequency distribution $g(\omega)=g(-\omega)$. 
Thus the order parameter $r$ is computed only in terms of 
{\it locked} oscillators as   
\begin{equation}
r=\sum_{k=1}^{\infty} p_d (k) 
\int_{-\frac{Jk\bar{r}}{\langle k\rangle}}^{\frac{Jk\bar{r}}{\langle k\rangle}} 
d\omega g(\omega) \int_0^{2\pi}   d\phi \, e^{i\phi} \, 
\delta\left[\phi-\sin^{-1}\left(\frac{\omega\langle k\rangle}{Jk\bar{r}}\right)\right]. 
\label{eq:rbarr}
\end{equation}
The effective field $\bar{r}$ is obtained by solving the following 
self-consistent equation derived from its definition,  
\begin{equation}
\bar{r}=
\sum_{k=1}^{\infty} {k p_d (k) \over \langle k \rangle }
	\int_{-\frac{Jk\bar{r}}{\langle k\rangle}}^{\frac{Jk\bar{r}}{\langle k\rangle}} 
	d\omega g(\omega) \int_0^{2\pi}   d\phi\,
	e^{i\phi} \delta\left[\phi-\sin^{-1}\left(\frac{\omega\langle k\rangle}{Jk\bar{r}}\right)\right]. 
\label{eq:sc}
\end{equation}
To solve Eqs.~(\ref{eq:rbarr}) and (\ref{eq:sc}), 
we expand them in powers of $\bar{r}$ as 
\begin{equation}
r=\sum_{n=1}^{\infty} a_{2n-1} 
\left(\frac{J}{\langle k\rangle}\bar{r}\right)^{2n-1} {\rm and}\  
\bar{r}=\sum_{n=1}^{\infty} \bar{a}_{2n-1} 
\left(\frac{J}{\langle k\rangle}\bar{r}\right)^{2n-1},
\label{eq:expand1}
\end{equation}
which are valid for finite $N$.
Here the coefficients are given by 
$a_{2n-1}=\langle k^{2n-1}\rangle \, u_{2n-1}$ and 
$\bar{a}_{2n-1}=(\langle k^{2n}\rangle/\langle k\rangle) \, u_{2n-1}$, where 
$\langle k^m \rangle=\sum_k k^m p_d(k)$, and   
\begin{equation}
u_{2n-1}=
\frac{(-1)^{n-1} 2^{-n-1/2} (n-3/2)!}{n!  (n-1)!},
\label{eq:coeff}
\end{equation}
from the relation 
$g(\omega)=(2\pi)^{-1/2}\sum_{n=0}^\infty  (-\omega^2/2)^n/n!$ and 
$\int_{-\pi/2}^{\pi/2} d\phi \, \cos^2\phi \, \sin^{2n-2}\phi = 
\pi^{1/2}(n-3/2)!/(2n!)$. 
Eq.~(\ref{eq:expand1}) is  valid also 
in the thermodynamic limit unless $a_{2n-1}$ or $\bar{a}_{2n-1}$ diverge.  
In case of SF networks with $p_d(k)\sim k^{-\lambda}$, however,
$\langle k^m\rangle$ diverges as $N^{m/(\lambda-1)-1}/(m-\lambda+1)$ 
for $m>\lambda-1$~\cite{lee04} 
and thus $a_{2n-1}$ $(n>\lambda/2)$ and 
$\bar{a}_{2n-1}$ $(n>(\lambda-1)/2)$ diverge:  
$a_{2n-1}\simeq c_{2n-1} N^{(2n-1)/(\lambda-1)-1}/(2n-\lambda)$ 
for $n>\lambda/2$ 
and $\bar{a}_{2n-1}\simeq \bar{c}_{2n-1} N^{(2n)/(\lambda-1)-1}/(2n-\lambda+1)$ 
for $n>(\lambda-1)/2$ with $c_{2n-1}$ and $\bar{c}_{2n-1}$ constants.
In terms of a scaling variable $q\equiv N^{1/(\lambda-1)} 
J\bar{r}/\langle k\rangle$, 
the terms with such diverging coefficients in 
Eq.~(\ref{eq:expand1}) are arranged as 
$(J\,\bar{r}/\langle k\rangle)^{\lambda-1} \sum_{n>\lambda/2} c_{2n-1} 
q^{2n-\lambda}/(2n-\lambda)=(J\,\bar{r}/\langle k\rangle)^{\lambda-1}\,
f_\lambda(q)$ for $r$ and 
$(J\,\bar{r}/\langle k\rangle)^{\lambda-2} \sum_{n>(\lambda-1)/2} 
\bar{c}_{2n-1} q^{2n-\lambda+1}/(2n-\lambda+1) = 
(J\,\bar{r}/\langle k\rangle)^{\lambda-2} \bar{f}_\lambda(q)$ 
for $\bar{r}$. The alternating sign and fast decay of $c_{2n-1}$ 
and $\bar{c}_{2n-1}$ from the behavior of $u_{2n-1}$ in Eq.~(\ref{eq:coeff})  
  make $f_\lambda(q)$ and $\bar{f}_\lambda(q)$ finite 
in the limit $q\to\infty$ that corresponds to $N\to\infty$ and 
$\bar{r}$ finite.
Consequently, the following equations should be considered 
for SF networks in the thermodynamic limit, 
\begin{eqnarray}
r&=&\sum_{n=1}^{\infty} a^{'}_{2n-1} 
\left(\frac{J}{\langle k\rangle}\bar{r}\right)^{2n-1} + 
a^{'}_{\lambda-1} \left(\frac{J}{\langle k\rangle}\bar{r}\right)^{\lambda-1}  
+\cdots,
\nonumber\\
\bar{r}&=&\sum_{n=1}^{\infty} \bar{a}^{'}_{2n-1} 
\left(\frac{J}{\langle k\rangle}\bar{r}\right)^{2n-1} + 
\bar{a}^{'}_{\lambda-2} 
\left(\frac{J}{\langle k\rangle}\bar{r}\right)^{\lambda-2} 
+\cdots,
\label{eq:expand2}
\end{eqnarray}
where all the coefficients are finite and in particular, 
$a^{'}_{2n-1} = a_{2n-1}$ for $n\leq \lfloor \lambda/2 \rfloor$ 
and $\bar{a}^{'}_{2n-1} = \bar{a}_{2n-1}$ for $n\leq \lfloor (\lambda-1)/2 
\rfloor$ with $\lfloor x\rfloor$ denoting here the greatest integer 
that is less than $x$. Notice that $a^{'}_{\lambda-1} 
= f_\lambda(\infty)$ and 
$\bar{a}^{'}_{\lambda-2}=\bar{f}_\lambda(\infty)$~\cite{coeff}.

The existence of the synchronization transition for $\lambda>3$ 
is identified by the emergence of non-zero values 
of  $r$ and  $\bar{r}$ 
only when $\bar{a}^{'}_1 J/\langle k\rangle>1$ or  $J>J_c$ with  
\begin{equation}
J_{\rm c}= 
2\sqrt{\frac{2}{\pi}}{\langle k \rangle^2 \over \langle k^2 \rangle}.
\label{eq:jc}
\end{equation}
Inspecting Eq.~(\ref{eq:expand2}) for $0<\Delta\equiv J/J_c-1\ll 1$, 
one finds that 
$r\simeq (\langle k\rangle^2/\langle k^2\rangle)\bar{r}$ and 
that $\Delta \bar{r}\simeq  
|\bar{a}^{'}_3| (J_c\bar{r}/\langle k\rangle)^3$ for $\lambda>5$ and 
$\Delta \bar{r} \simeq  
|\bar{a}^{'}_{\lambda-2}| (J_c\bar{r}/\langle k\rangle)^{\lambda-2}$ 
for $3<\lambda<5$. 
Thus it holds for small positive $\Delta$ and $\lambda>3$ that 
\begin{equation}
r\sim \Delta^\beta,
\label{eq:r1}
\end{equation}
with the order-parameter critical exponent $\beta$ given by 
\begin{equation}
\beta=\left\{
\begin{array}{ll}
\frac{1}{2} & {\rm for}~\lambda>5,\\
\frac{1}{\lambda-3}&{\rm for}~3<\lambda<5.
\end{array}
\right.
\label{eq:beta}
\end{equation}
Notice that in the globally-coupled case 
where $\langle k^2\rangle = \langle k\rangle^2 =(N-1)^2$, 
the critical point $J_c$ is $2\sqrt{2/\pi}\simeq 1.60$ and the 
critical exponent $\beta$ is $1/2$~\cite{kuramoto,strogatz00}.
In Ref.~\cite{moreno04}, the exponent $\beta$  
has been shown numerically to be close to $1/2$  
in the synchronization transition on 
the Barab\'{a}si-Albert network~\cite{barabasi99} that has $\lambda = 3$. 

When $2<\lambda<3$, $\langle k^2\rangle$ is  
$\mathcal{O}(N^{(3-\lambda)/(\lambda-1)})$~\cite{lee04} and 
thus the critical point $J_c$ 
is of order $N^{-(3-\lambda)/(\lambda-1)}$,  
which is  zero in the thermodynamic limit. 
It means that $\bar{r}$ and $r$ 
are always non-zero  values for non-zero values of $J$. 
Eq.~(\ref{eq:expand2}) is arranged into $r\simeq 2^{-3/2}\pi^{1/2} 
J \bar{r}$ and 
$\bar{r}\simeq |\bar{a}^{'}_{\lambda-2}|
(J\bar{r}/\langle k\rangle)^{\lambda-2}$ for small $J$, which leads to 
\begin{equation}
\bar{r}\sim J^{\frac{\lambda-2}{3-\lambda}} \quad {\rm and}\quad
r\sim J^{\frac{1}{3-\lambda}}. 
\label{eq:r2}
\end{equation}

\begin{figure}
\includegraphics[width=\columnwidth]{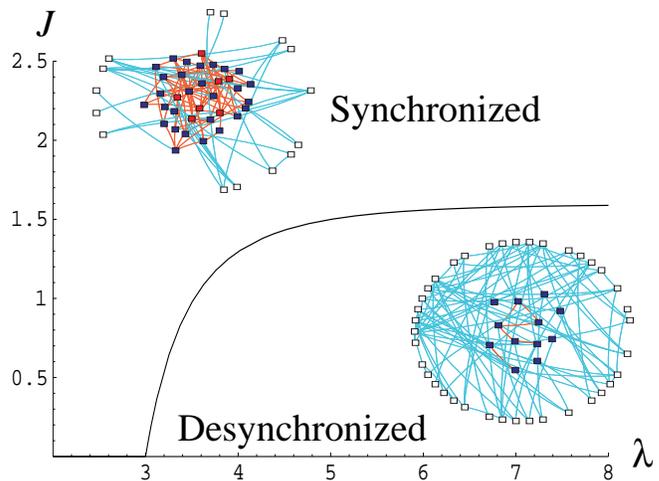}
\caption{ 
(Color online) Phase diagram and typical dynamical states of oscillators on 
SF networks. 
The phase boundary in Eq.~(\ref{eq:jc}) is drawn with 
$p_d(k)=k^{-\lambda}/\zeta(\lambda)$ for $k\geq 1$ and 
$\zeta(\lambda)$ the Riemann-zeta function. 
The two network configurations  represent dynamical states of 
coupled oscillators  
that have evolved with $J=1.0$ (desynchronized) and $J=4.0$ 
(synchronized) respectively on SF networks  with $\lambda=3.6$. 
In both, synchronized nodes, those  having their phases $\phi_i$ 
so close to the average phase $\theta$ 
that $|\phi_i-\theta|<0.5$,   
are drawn as filled boxes inside the circle    
while the others are as empty boxes at the circumference. 
Edges connecting synchronized nodes are drawn as thick lines. 
}
\label{fig:pd}
\end{figure}

\begin{figure*}
\includegraphics[width=0.66\columnwidth]{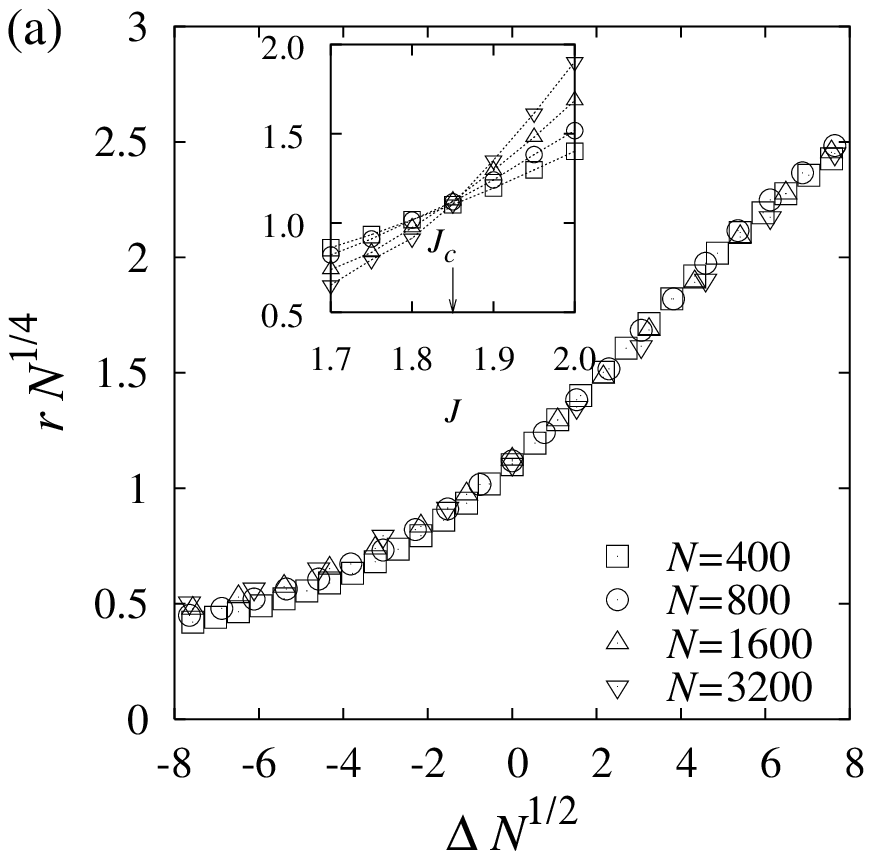}
\includegraphics[width=0.66\columnwidth]{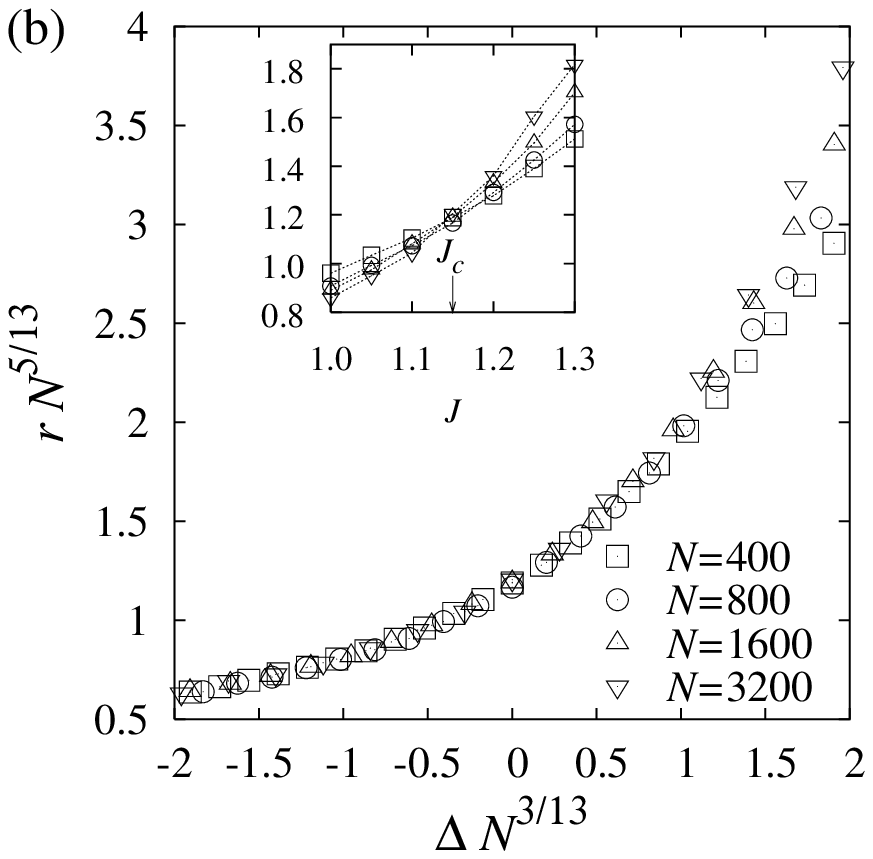}
\includegraphics[width=0.66\columnwidth]{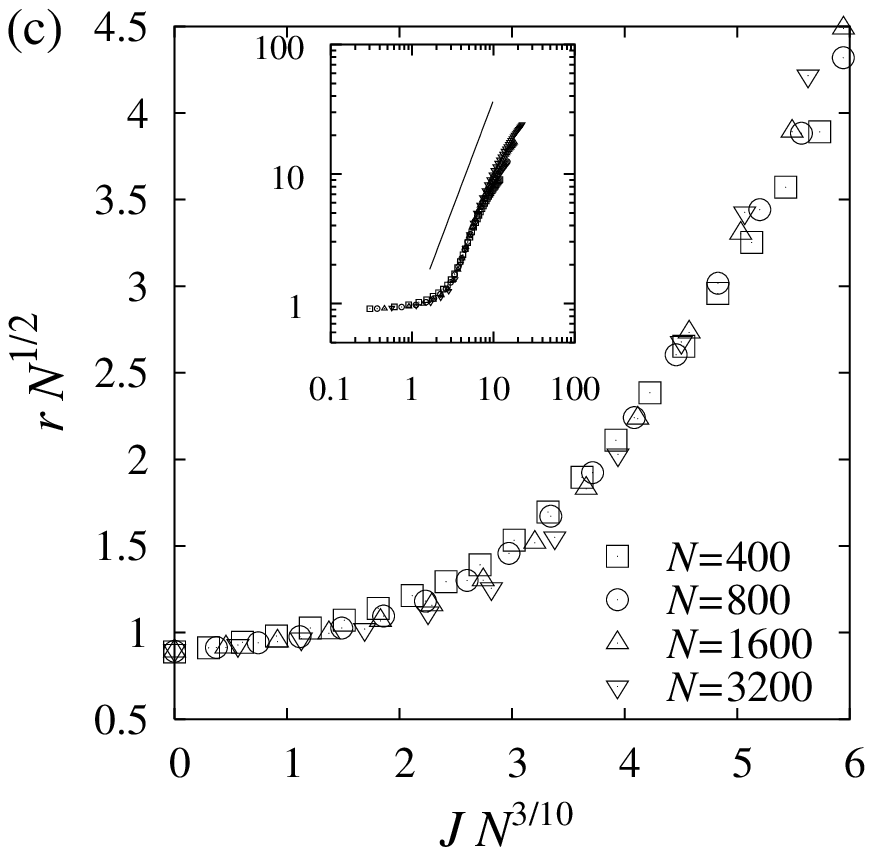}
\caption{Data collapse of scaled order parameter 
for $\langle k\rangle=4$, (a) $\lambda=6.4$, 
(b) $\lambda=3.6$, and (c) $\lambda=2.4$ according to 
Eqs.~(\ref{eq:scal1}) and (\ref{eq:scal2}). 
In (a) and (b), insets show the crossing of 
scaled order parameters at $J_c$, which is found to be 
(a) $1.85(5)$ and (b) $1.15(5)$, larger than $1.22$ and 
$0.84$, respectively, evaluated from Eq.~(\ref{eq:jc}). In the inset of (c), 
asymptotic behavior of the scaling function is explicitly 
shown. The solid line has  slope $1/(3-\lambda)=5/3$.}
\label{fig:fss}
\end{figure*}

The phase diagram and dynamical states of the oscillators  
in the desynchronized state 
and the synchronized state are shown in  Fig.~\ref{fig:pd}. 
The only contribution to the order parameter in the desynchronized state 
is made by temporary synchronization of a few oscillators that 
could be disconnected.
On the contrary, a finite fraction of oscillators have their phases 
close to the average phase $\theta(t)$ in the synchronized state. 
This giant cluster of synchronized nodes contains a finite fraction of 
edges as well, implying 
that the synchrony originates from the interaction between 
them through the SF coupling topology.
Furthermore, the continuously-varying critical exponent $\beta$
according to the degree exponent $\lambda$ demonstrates that  
the variation of the network topology may bring about 
the change of the universality class of the synchronization 
transition on complex networks. 
The influence of the structural organization on the emergence 
of  dynamical order is more significant in case of $2<\lambda<3$ 
where $J_c=0$.
A deeper understanding of the nature of the synchronization 
transition is available when we investigate
the finite-size scaling behavior of the order parameter.
As we shall see, it illuminates different nature of 
the synchronization transition 
between SF networks with $2<\lambda<3$ and $\lambda>3$ 
as well as allows  
to confirm our findings, Eqs.~(\ref{eq:jc}), (\ref{eq:r1}), 
(\ref{eq:beta}), and (\ref{eq:r2}),  numerically. 

\section{Derivation of finite-size scaling behavior}
\label{sec:fss}
In this section, we describe how to compute the order parameter 
in the critical regime and combining it with the results in the previous 
section, present the finite-size scaling behavior 
of the order parameter. 

In finite-size systems, the order parameter $r$ is non-zero even with $J=0$ 
so that $r_{J=0}\sim N^{-1/2}$, as  
the phases are distributed uniformly in $[0,2\pi)$.
With $J$ increasing from $J=0$,
the clusters,  connected oscillators with nearly the same phases, 
form and evolve to find 
the largest one among them  
much larger than $r_{J=0}N\sim N^{1/2}$, when the 
order parameter $r$ can be approximated by 
$r\simeq S/N$
with $S$ the (ensemble-averaged) largest cluster size.
The synchronized state is marked by the 
presence of a giant cluster whose size is $\mathcal{O}(N)$ 
making $r$ finite. 

To understand the jump to ${\cal O}(1)$ of the 
order parameter in the critical regime 
for $\lambda>3$ and in the small $J$ regime for $2<\lambda<3$,  
we should know the largest cluster size $S_c$ in those regimes.  
$S_c$ is of order $N^\alpha$ with $1/2\leq\alpha<1$, and we will 
try to extract the exponent $\alpha$ from the asymptotic behavior 
of the cluster size distribution, which can be obtained using 
Eq.~(\ref{eq:expand2}), but only for $\lambda>3$.
For $2<\lambda<3$, we will see that the largest cluster size 
increases gradually from $\mathcal{O}(N^{1/2})$ to 
$\mathcal{O}(N)$ without any characteristic size between them 
while $J$ is small.

Let us  define  the generating functions $\mathcal{P}(z)=\sum_s 
P(s) z^s$ and 
$\mathcal{\bar{P}}(z)=\sum_s \bar{P}(s) z^s$, where $P(s)$  
is the probability that a node belongs to a size-$s$ cluster 
and $\bar{P}(s)$ the probability that an edge  
is followed by a size-$s$ cluster.  
The dependence on the system size $N$ is implicitly included.
The generating functions are analytic for finite $N$ and thus 
one can expand the inverse function $\mathcal{\bar{P}}^{-1}(\omega)$ as 
$z = \mathcal{\bar{P}}^{-1}(\omega)=1-\sum_{n\geq 1} b_n (1-\omega)^n$ 
around $\omega=1$.
When $z=z^*_N\equiv e^{-\frac{1}{\tilde{S}}}$ with $\tilde{S}$ satisfying 
$S_2\ll \tilde{S}\ll S$ and $S_2$ being the second largest cluster size, 
the corresponding value $\omega^*_N = \mathcal{\bar{P}}(z^*_N)$ 
represents the statistical weight of all clusters but the 
largest one, that is, $\omega^*_N\simeq\sum_{s<S} \bar{P}(s)$.
Suppose that Eq.~(\ref{eq:expand2}) is valid in the critical regime. 
Since $\bar{r}=\lim_{N\to\infty} (1-\omega^*_N)$,
one can see that 
the coefficients $b_n$ should take such values as 
enable the equation $z=1-\sum_{n\geq 1} b_n (1-\omega)^n$ with 
$z=\lim_{N\to\infty} z^*_N=1$ to be reduced to Eq.~(\ref{eq:expand2}) 
with $\bar{r}$ replaced by $1-\omega$.  
Consequently we have $z = \omega + \sum_{n=1}^\infty
\bar{a}^{'}_{2n-1}[J(1-\omega)/\langle k\rangle]^{2n-1} + 
\bar{a}^{'}_{\lambda-2} 
[J(1-\omega)/\langle k\rangle]^{\lambda-2}+\cdots$ with 
$\omega=\mathcal{\bar{P}}(z)$.

The last relation informs us that   
at the critical point 
where $\bar{a}^{'}_1J/\langle k\rangle=1$, 
$1-\mathcal{\bar{P}} (z)$  has the following leading terms: 
$(1-z)^{1/3}$ for $\lambda>5$, 
$(1-z)^{1/(\lambda-2)}$ for $3<\lambda<5$, and 
$J^{\lambda-2}(1-z)^{\lambda-2}$ for $2<\lambda<3$, respectively. 
Eq.~(\ref{eq:expand2}) also enables us to know that $\mathcal{P}(z)$ is related 
to $\mathcal{\bar{P}}(z)$ via 
$1-\mathcal{P}(z)\sim J(1-\mathcal{\bar{P}}(z))$ and thus 
the leading terms of $1-\mathcal{P}(z)$  are given as 
$(1-z)^{1/3}$ for $\lambda>5$, 
$(1-z)^{1/(\lambda-2)}$ for $3<\lambda<5$, and 
$J^{\lambda-1}(1-z)^{\lambda-2}$ for $2<\lambda<3$, respectively. 
Here we showed the $J$ dependence explicitly only for $2<\lambda<3$.
These singularities give 
the asymptotic behaviors of $P(s)$ through  the relations 
$(1/s!)(d^s/dz^s) {\cal P}(z)|_{z=0} = P(s)$ and  
$(1/s!)(d^s/dz^s) (1-z)^{\tau-1}|_{z=0} \sim s^{-\tau}$ for 
large $s$. 
Finally the largest cluster size $S_c$ 
at the critical point can be obtained using the relation 
$\sum_{s<S_c} P(s) = 1- S_c/N$~\cite{cohen02}. 
The result is    
\begin{equation}
S_c\sim 
\left\{
\begin{array}{ll}
N^{\frac{3}{4}} &{\rm for} ~ \lambda>5,\\
N^{\frac{\lambda-2}{\lambda-1}}&{\rm for}~ 2<\lambda<5. \\
\end{array}
\right.
\label{eq:rc}
\end{equation}

For $\lambda>3$, 
$S_c$ in Eq.~(\ref{eq:rc}) is much larger than 
$r_{J=0}N\sim N^{1/2}$ and we can accept it as true.
The oscillators are not ready to form the giant cluster 
until the coupling strength reaches the critical value. 
There emerge clusters of 
all sizes up to $S_c$ at the critical point 
and many of them merge into the giant cluster with the 
coupling strength slightly increased.
Combining Eqs.~(\ref{eq:r1}) and  (\ref{eq:rc}), and  
introducing another exponent $\mu$ defined by 
$S_c/N\sim N^{-\beta/\mu}$, which is given as
\begin{equation}
\mu=\left\{
\begin{array}{ll}
2 & {\rm for}~\lambda>5,\\
\frac{\lambda-1}{\lambda-3}&{\rm for}~3<\lambda<5,
\end{array}
\right.
\label{eq:mu}
\end{equation}
we obtain the following finite-size scaling behavior of $r$ 
\begin{equation}
r=N^{-\beta/\mu}\, \Psi(\Delta N^{1/\mu}). 
\label{eq:scal1}
\end{equation}
The scaling function $\Psi(x)$ is a constant for $x\ll 1$ while 
behaves as $x^\beta$ for $x\to\infty$.
We remark that $\mathcal{O}(N^{-1/4})$ fluctuation of $r$ at $J=J_c$ 
for $\lambda>5$ has also been found 
in the Kuramoto model~\cite{daido} 
as well as in SW networks~\cite{hong02}.
We plot the scaled numerical data for $\lambda=6.4$ and $3.6$ in 
Fig.~\ref{fig:fss} (a) and (b), respectively, 
which show good agreement with Eq.~(\ref{eq:scal1}) 
except for slight deviations of $J_c$ from Eq.~(\ref{eq:jc}).
Such deviations of the critical point  have also been observed 
in the Monte Carlo simulation of the Ising model on 
SF networks~\cite{herrero04}, but are not 
so serious as to give rise to a finite value of $J_c$ even for 
$2<\lambda<3$. 
It is known that 
the critical point can be more precisely predicted by 
studying the system on the Cayley tree with a given degree 
distribution~\cite{dorogovtsev02}.

When $2<\lambda<3$, $S_c$ in Eq.~(\ref{eq:rc}) is much 
smaller than $r_{J=0}N$. This means that  
Eq.~(\ref{eq:expand2}) is not correct in the critical regime 
because of strong finite-size effects.
Instead, 
the finite-size scaling behavior of the order parameter for $2<\lambda<3$ 
can be obtained by connecting 
$r\sim J^{1/(3-\lambda)}$ in Eq.~(\ref{eq:r2})  and 
$r_{J=0}\sim N^{-1/2}$, which leads to 
\begin{equation}
r=N^{-\frac{1}{2}} \Phi(JN^{\frac{3-\lambda}{2}}),
\label{eq:scal2}
\end{equation}
where $\Phi(x)$ is a constant for $x\ll 1$ 
and $\Phi(x)\sim x^{1/(3-\lambda)}$ for $x\to\infty$.
This behavior is confirmed by 
the numerical data with $\lambda=2.4$ in Fig.~\ref{fig:fss} (c).

Distinctive finite-size scaling behaviors shown in Eqs.~(\ref{eq:scal1}) 
and (\ref{eq:scal2}) represent a significant difference in the emergence 
of  dynamical order between SF networks with $\lambda>3$ and 
$2<\lambda<3$. 
In SF networks with $2<\lambda<3$,  
the giant cluster of synchrony is formed if only $J\gg N^{-(3-\lambda)/2}$. 
A large number of hub nodes that are connected with one another 
do not allow separate clusters of various sizes to 
be developed but force most nodes to belong to the unique giant cluster 
for any non-zero coupling strength. 
It is completely different from the way of forming the giant cluster 
in SF networks with $\lambda>3$.  Clusters of various sizes 
up to $\mathcal{O}(N^{1-\beta/\mu})$ in the critical regime
merge to form the giant cluster  as the system  
escapes from the critical regime.
The advantage of many real complex systems having their 
degree exponents between $2$ and $3$ in their network structures 
is thus very obvious. Such  highly heterogeneous systems 
can remain in the dynamically-ordered state only with a very small 
coupling strength, e.g., $\mathcal{O}(N^{-(3-\lambda)/2})$ in 
the synchronization phenomena, which is required to maintain 
the stability of the system. 
On the other hand, a perturbation can propagate easily 
through the hubs as revealed by the linear stability analysis of 
the ordered state~\cite{barahona02,nishikawa02,hong04}, 
necessary to adapt the system to the changes of the environment.

\section{Summary and discussion}
\label{sec:summary}
We investigated the synchronization transition of 
coupled limit-cycle oscillators on complex networks 
analytically and numerically.  
We used the mean-field approach to find the 
critical coupling strength and the critical exponent. 
Furthermore, we showed how the self-consistent equations
derived by the mean-field approach 
can be used also to compute the critical fluctuation of the order parameter 
that completes the finite-size scaling analysis. 
We believe that this method can be generally applied to the study of 
finite-size effects in complex networks.
The synchronization transition turned out to be 
closely related to the formation of the giant cluster of 
synchronized oscillators, 
which crucially depends on the connectivity pattern of a given 
network such that the critical phenomena are distinguished 
according to the degree exponent. 
Our findings imply that complex systems take very heterogeneous 
connectivity patterns to acquire both stability and flexibility 
with only a small interaction among the elements.
While preparing this manuscript, we have learned of the 
work by T.~Ichinomiya~\cite{ichinomiya}, 
which partly overlaps with ours.

\begin{acknowledgements}
We are indebted to the network group at SNU, where this work 
was initiated, for helpful discussions. We thank 
Byungnam Kahng and Heiko Rieger for comments on the manuscript. 
This work was supported by the Deutsche Forschungsgemeinschaft (DFG).
\end{acknowledgements}

\end{document}